\documentclass[12pt]{article}

\usepackage{bbold}
\usepackage{amssymb}
\usepackage{slashed}
\usepackage{graphics}

\declareslashed{}{/}{.1}{0}{E}

\usepackage{times}

\newcommand{\sy}[1]{(\hspace{-.01mm}#1)}
\newcommand{\syt}[1]{(\hspace{-.01mm}#1)^{\!\rm T}}
\newcommand{\sytt}[1]{(\hspace{-.01mm}#1)^{\!\rm T\!T}}

\def\D{{\cal D}}

\def\r{\rho}
\def\s{\sigma}

\def\Sl#1{\slashed{#1}}

\def\wt#1{\widetilde{#1}}

\def\ss{\scriptstyle}
\def\sss{\scriptscriptstyle}

\def\m{\mu}
\def\n{\nu}

\def\be{\begin{equation}}
\def\ee{\end{equation}}
\def\beq{\begin{equation}}
\def\eeq{\end{equation}}
\def\bea{\begin{eqnarray}}
\def\eea{\end{eqnarray}} 
\def\beqa{\begin{equation}\begin{array}{l}}
\def\eeqa{\end{array}\end{equation}}

\def\eqn#1{(\ref{#1})}

\def\eqref#1{eq.~(\ref{eq:#1})}

\def\a{\alpha}
  \def\g{\gamma}

\def\L{{\it\Lambda}}

\def\nn{\nonumber}

\begin{document}

\thispagestyle{empty}
\begin{flushright}
\framebox{\small BRX-TH~487
}\\
\end{flushright}

\vspace{.8cm}
\setcounter{footnote}{0}
\begin{center}
{\Large{\bf 
Gauge Invariances and Phases of Massive Higher Spins in (A)dS
}
    }\\[10mm]

{\sc S. Deser
and A. Waldron
\\[6mm]}

{\em\small  
Physics Department, Brandeis University, Waltham,
MA 02454, 
USA\\ {\tt deser,wally@brandeis.edu}}\\[5mm]

\bigskip

\bigskip

{\sc Abstract}\\
\end{center}

{\small
\begin{quote}

The $(m^2,\L)$ plane of spin $s>1$ massive fields in (A)dS
backgrounds is shown to consist of separate phases, divided by lines of 
novel ``partially massless'' gauge theories that 
successively remove helicities, starting from the lowest, 0 or $\pm1/2$.
The norms
of the excluded states flip as the
gauge lines are crossed and only the 
region containing the 
massive Minkowski theory is unitary. 
The partially massless gauge theories are unitary or not, depending
on the ordering of the gauge lines.
This ``level splitting'' of
massless Minkowski gauge theories is specific to non-zero $\L$.

\bigskip

\bigskip

\end{quote}
}

\newpage






Gauge invariance, masslessness, null propagation and helicity $\pm s$
excitations are all synonymous in flat space, a degeneracy
that is lifted in (A)dS. 
We show that, for massive spin $s>1$ theories, 
new gauge invariances allow  ``partially massless'' propagation. 
Specifically, 
there are intermediate theories
with $2,4,\ldots ,2s,2s+1$ propagating degrees of freedom (PDoF) 
for bosons or
$2,4,\ldots, 2s$ PDoF for fermions\footnote{The term helicity 
is used in a correspondence sense with Minkowski space. 
Also, a new feature for $s\geq5/2$
is that auxiliary fields are necessary:
in the massless limit they propagate but decouple. In passing, we note
that null propagation is generically also lost in (A)dS~\cite{Deser:1983tm}}.

The physics of massive higher spin fields in (A)dS
is best displayed in the $(m^2,\L)$-plane 
where, phases describing $2s+1$ massive PDoF
are separated by lines where gauge invariances remove subsets of 
lowest helicity modes.
Further, the norms of the excised helicities
flip sign as these gauge lines are crossed. As a consequence:
(i) Only the phase occupied by the flat
massive theory is unitary, since the set of excitations flipping sign
is distinct on each line. (ii) Unitarity of the partially massless theories 
requires that their gauge invariances remove any negative norm states. 
This occurs only 
when, starting from the
unitary Minkowski region, the 
line removing the lowest helicity state(s) can be traversed
first. Subsequent lines must also be be traversed in order,
ending on the strictly massless helicity $\pm s$ line.

To see how massive constraints are transmuted
to Bianchi identities along lines in the $(m^2,\Lambda)$ plane, consider
first the $s=1$ (A)dS Proca equation
${\cal G}_\m\equiv(D^2-m^2-\L)\phi_\m-D_\m D.\phi=0$
whose divergence
$D.{\cal G}=-m^2 D.\phi$
implies $3=2s+1$ PDoF. Along the line $m^2=0$, the
constraint becomes a Bianchi identity
reflecting the gauge invariance $\delta\phi_\m=D_\m\xi$
and the system describes helicities $\pm1$ only. 
For $s>1$, however, two new things happen: (i) The field equations have 
$[s]$ open indices, so each of $[s]$ successive
divergences yield constraints implying $2s+1$ PDoF.
(ii) The monomial $m^2$ in the Proca constraint is generalized to 
non-trivial functions of $(m,\L)$
whose vanishing yields new Bianchi identities, an effect  
that clearly degenerates at  $\L=0$.

Each Bianchi identity implies gauge invariances which remove  
corresponding lower helicities from the spectrum. 
In addition, the coefficients of the divergence
constraints flip sign across the gauge boundaries
and appear as denominators 
in canonical (anti)commutators.
Hence, whenever gauge invariance removes a particular helicity,
its norm has opposite signs in the adjacent massive phases and 
the $(m^2,\L)$-plane is partitioned into unitarily allowed and forbidden
regions\footnote{ 
The constant curvature background does not suffer
the usual gravitational coupling 
problems in the unitarily allowed regions, although the
old causality difficulties  
of $s=3/2$ in E/M backgrounds~\cite{Johnson:1961vt,Velo:1969bt} 
resurface for the forbidden regions.}.

In this Letter, we explicitly treat spins
$s=3/2$, $2$, $5/2$ and $3$. It is known
that, beyond its vector gauge invariance at
$m^2=0$, spin~2 is partially massless along the line
$m^2=2\L/3$ thanks to
a Weyl-like scalar gauge invariance and accompanying 
double divergence Bianchi
identity~\cite{Deser:1983tm}.
The dS region $m^2<2\L/3$ is unitarily forbidden~\cite{Higuchi:1987py}.
We extend this result to spin~3/2 and then show that spins~5/2 and~3
possess new partially massless gauge lines.
Finally we present a PDoF counting argument indicating that partial
masslessness is enjoyed by all spins $s>3/2$ in (A)dS.



The massive spin~3/2 field equation 
is\footnote{All our field equations are derived
from hermitean actions.}
\be
{\cal R}_\m\equiv\g_{\m\n\r}\D^\n\psi^\r=0\; ,
\ee
\be
\D_\m\equiv
D_\m+\frac{1}{2}\,m\,\g_\m\ , 
\qquad [D_\m,D_\n]\,\psi_\r= \frac{2\L}{3}\,g_{\r[\m}\psi_{\n]}
+\frac{\L}{6}\,\g_{\m\n}\psi_\r\, .
\label{1}
\ee
To test for Bianchi identities, compute the ``curly'' divergence of~\eqn{1}
\be
\D.{\cal R}=-\frac{3}{2}\,(m^2+\L/3)\,\g.\psi\, ;
\label{wbb}
\ee
this is obviously a constraint
whose coefficient vanishes along the gauge boundary $m^2=-\L/3$
familiar from cosmological SUGRA~\cite{Townsend:1977qa}. Along this line 
the theory is gauge invariant under variations 
$\delta\psi_\m=\D_\m\varepsilon$ and describes
only helicities $\pm3/2$ in AdS ($\L<0$). 

The appearance of a ghost across this line 
can be detected by examining
the 0-components of the equal time anticommutator\footnote{This method
was originally used to exhibit the inconsistency of
spin~3/2 E/M couplings~\cite{Johnson:1961vt}.}
\bea
\{\psi_0(t,\vec x),\psi^\dagger_0(t,\vec x')\}&=&
\frac{\cosh^{-2}(Mt)\,(-\,^{(3)}\!D^2-\L/4)}{3\,(m^2+\L/3)}
\;\frac{1}{\sqrt{-g}}\,\delta^3(\vec x-\vec x')\;,\nn\\
\label{ac}
\eea
in the synchronous metric 
$
ds^2=-dt^2+\cosh^2(Mt)\,(dr^2$ $+\sin^2(Mr)\,(d\theta^2$
$+$ $\sin^2\theta\,d\phi^2))$,
$M^2\equiv\L/3$, that  describes dS, and also AdS when continued to
imaginary $M$.
Unitarity requires positivity of the operator on
the r.h.s. of~\eqn{ac}. For AdS (ignoring spatial boundary
terms) 
the intrinsic 3-Laplacian $-{}^{(3)}\!D^2$ and $-\L/4$ are separately positive.
In dS, the lowest eigenvalue of $-{}^{(3)}\!D^2$ acting
on spinors is $\L/4$ (the spatial slices are an $S^3$ there) 
so $-{}^{(3)}\!D^2-\L/4\geq0$.
Therefore, the coefficient $m^2+\L/3$ 
in the would-be Bianchi identity~\eqn{wbb}
controls the sign of the anticommutator~\eqn{ac} and only the region
$m^2\geq-\L/3$ is unitary\footnote{As a consequence
the $m\rightarrow0$ limit can be taken only in dS, contrary 
to~\cite{Grassi:2001dm}. As discussed in~\cite{Deser:2000de}, the
true massless limit is the AdS one,
$m\rightarrow\sqrt{-\L/3}$. This tuning of $(m,\L)$ also 
implies that unitary SUSY 
representations exist only in AdS~\cite{Pilch:1985aw}.
Unitary non-supersymmetric models in dS are, of
course, not ruled out.}; the phase diagram is
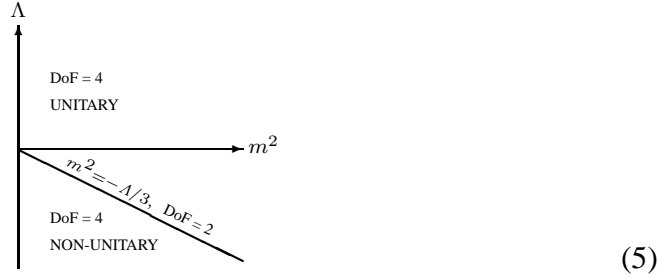
\be
\begin{picture}(300,100)(-100,5)
\put(0,100){$\ss \Lambda$}
\put(90,48.5){$\ss m^2$}


\put(20,44){\rotatebox{334}{$\sss m^2=-\L/3 ,\;\;\mbox{\tiny DoF = 2}$}}
\put(15,75){\tiny DoF = 4}
\put(15,65){\tiny UNITARY}
\put(15,22){\tiny DoF = 4}
\put(15,12){\tiny NON-UNITARY}



\put(3,50){\vector(1,0){85}}
\put(3,5){\vector(0,1){92}}

\thicklines

\put(3,50){\line(2,-1){85}}




\end{picture}
\ee

We emphasize that poles in canonical
(anti)commutators  at gauge  values of 
$(m^2,\Lambda)$ are a general feature of massive higher spins.


Spin~2 has 
two gauge lines, with phase diagram
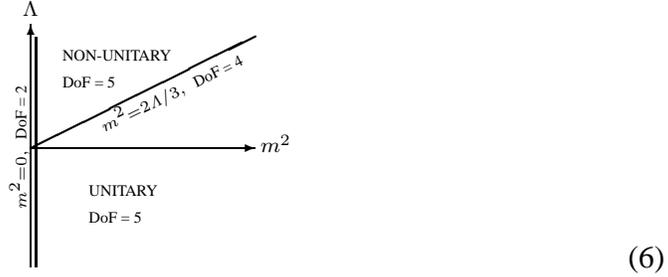
\be
\begin{picture}(300,100)(18,5)

\put(120,100){$\ss \Lambda$}
\put(210,48.5){$\ss m^2$}

\put(148,56){\rotatebox{26}{$\sss m^2=2\L/3 ,\;\;\mbox{\tiny DoF = 4}$}}
\put(115,27){\rotatebox{90}{$\sss m^2=0 ,\;\;\mbox{\tiny DoF = 2}$}}
\put(135,83){\tiny NON-UNITARY}
\put(135,73){\tiny DoF = 5}
\put(145,22){\tiny DoF = 5}
\put(145,32){\tiny UNITARY}

\put(123,50){\vector(1,0){85}}
\put(123,5){\vector(0,1){92}}

\thicklines

\put(125,5){\line(0,1){87}}
\put(123,50){\line(2,1){85}}

\end{picture}
\ee
Linearized cosmological Einstein gravity remains at $m^2=0$,
so the $(m^2,\L)$ half-plane is divided into two distinct
regions, just as for spin~3/2. The new partially massless gauge line 
at $m^2=2\L/3$ is the Weyl-like theory of~\cite{Deser:1983tm}. 
A detailed
canonical analysis proves that it describes 4
unitary PDoF~\cite{Deser:2001}
corresponding to the upper helicities $(\pm2,\pm1)$, whose
propagation is null~\cite{Deser:1983tm}. The dS region $m^2<2\L/3$
is unitarily forbidden~\cite{Higuchi:1987py} 
by an analysis completely equivalent
to that of spin~3/2 above.


The spin~5/2 field equations
are~\footnote{See~\cite{Singh:1974rc} for 
Minkowski higher spin massive field
equations. They can also be derived by
Kaluza--Klein reduction~\cite{Aragone:1987yx} from their simpler $d=5$
massless
antecedents~\cite{Fronsdal:1978rb}
and the
constraints then follow from the $d=5$ Bianchi identities~\cite{Deser:2001}.}
\bea
{\cal R}_{\m\n}&=&
\Sl D\,\psi_{\m\n}
+g_{\m\n}\,(\g.D.\psi-\frac{1}{2}\,\Sl D\,\psi_\r{}^\r)\nn\\
&+&\g_{(\m}\,(D_{\n)}\psi_\r{}^\r+2\Sl D \g.\psi_{\n)}-2D.\psi_{\n)})
-2D_{(\m}\g.\psi_{\n)}\nn\\
&+&m\,(\psi_{\m\n}-2\,\g_{(\m}\g.\psi_{\n)}-\frac{1}{2}\,g_{\m\n}\,
\psi_\r{}^\r)
-\frac{5}{12}\,\mu\,g_{\m\n}\,\chi\, ,
\label{huey}\\
{\cal R}_5&=&-\alpha\,(\Sl D-3m)\,\chi-\frac{5}{12}\,\mu\,\psi_\r{}^\r\, .
\label{duey}
\eea
where the spinor $\chi$ (with its equation of motion ${\cal R}_5$)
exemplifies the auxiliary fields needed for $s\geq5/2$ to impose
$2s+1$ massive PDoF. Its coupling $\mu$ is uniquely fixed,
$\mu^2=\frac{12\alpha}{5}\,(m^2+4\L/3)$, and it decouples 
on the strictly massless line $m^2=-4\L/3$.
The parameter $\alpha$ is free, modulo requiring reality of $\mu$ 
(and of the action). It 
can be set to $1$ when $m^2>-4\L/3$ and $-1$ for $m^2<-4\L/3$. This
discontinuity is a precursor of the non-unitary behavior we find
for the 
massive and partially massless theories in the latter region.

Remarkably, the field equations enjoy two distinct gauge 
invariances\footnote{A tilded index denotes its gamma-traceless part,
$X_{\wt \m}\equiv X_\m-(1/4)\,\g_\m\g.X$. We denote the
symmetric-traceless part of a pair indices by $X_{\{\m\n\}}\equiv X_{(\m\n)}
-(1/4)\,g_{\m\n} X_\r{}^\r$; we symmetrize with unit weight.},
\bea
\delta\psi_{\m\n}=D_{(\m}D_{\wt \n)}\,\varepsilon
+\frac{5\L}{16}\,g_{\m\n}\,\varepsilon\, ,\qquad&&
\delta\chi=-\frac{1}{8\a}\,\sqrt{15\a\L}\,(\Sl D+\sqrt{-3\L})
\,\varepsilon \quad\label{ra}\\
\delta\psi_{\m\n}=D_{(\m}\varepsilon_{\wt
\n)}+\frac{1}{2}\,\sqrt{\frac{-\!\L\,}{3}} \,\g_{(\m}\,\varepsilon_{\wt
\n)}\, ,&&\delta\chi=0\, ,
\label{ar}
\eea
along the respective AdS gauge lines
\be
m^2+\L/3=0\;,\quad m^2+4\L/3=0\, .\label{ladle} 
\ee
The vector-spinor invariance yields the
strictly massless theory, with helicities $\pm5/2$, whereas the novel spinor
invariance~\eqn{ra}
removes helicities $\pm1/2$ leaving
$(\pm5/2,\pm3/2)$.
Each invariance is accompanied by a Bianchi identity, which
turns into a constraint ensuring
$6=2s+1$ massive PDoF when 
$(m^2,\L)$ are detuned from their gauge values~\eqn{ladle}.

The coefficients $(m^2+4\L/3)$
and $(m^2+\L/3)$ control the positivity of equal time anticommutators.
Therefore, since the 
gauge lines all 
lie in AdS, the $(m^2,\L)$-plane
is divided into 3 regions, only the one including Minkowski space
$m^2>-4\L/3$ being unitary:
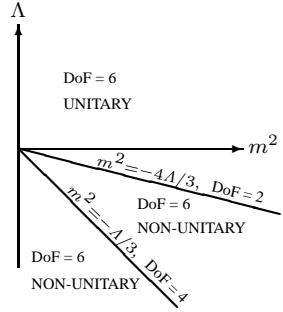
\be
\begin{picture}(300,120)(-100,-15)
\put(0,100){$\ss \Lambda$}
\put(90,48.5){$\ss m^2$}


\put(32,44){\rotatebox{347}{$\sss m^2=-4\L/3 ,\;\;\mbox{\tiny DoF = 2}$}}
\put(20,34){\rotatebox{316}{$\sss m^2=-\L/3 ,\;\;\mbox{\tiny DoF = 4}$}}
\put(20,75){\tiny DoF = 6}
\put(20,65){\tiny UNITARY}
\put(48,27){\tiny DoF = 6}
\put(48,18){\tiny NON-UNITARY}
\put(8,7){\tiny DoF = 6}
\put(8,-3){\tiny NON-UNITARY}

\put(3,50){\vector(1,0){85}}
\put(3,5){\vector(0,1){92}}

\thicklines

\put(3,50){\line(4,-1){99}}
\put(3,50){\line(1,-1){60}}

\end{picture}
\ee
Although the strictly massless AdS $m^2=-4\L/3$ theory is unitary, the
partially massless one is not, as
it fails the line ordering requirement: Starting from the unitary Minkowski
region where all norms are positive,
one would like first to traverse the line $m^2=-\L/3$, but that is only
possible in dS with negative $m^2$ (imaginary values of $m$ violate
hermiticity of the action and unitary evolution). Crossing the AdS
strictly massless line $m^2=-4\L/3$ first flips the norm of helicities
$(\pm3/2,\pm 1/2)$ so the partially massless AdS theory cannot be unitary.


Spin~3 enjoys new 
unitary partially massless gauge lines: 
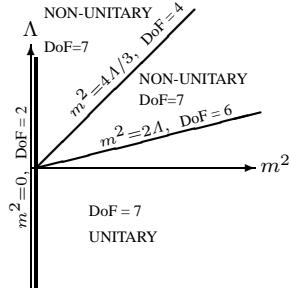
\be
\begin{picture}(300,100)(138,10)

\put(240,100){$\ss \Lambda$}
\put(330,48.5){$\ss m^2$}

\put(235,27){\rotatebox{90}{$\sss m^2=0 ,\;\;\mbox{\tiny DoF = 2}$}}
\put(255,69){\rotatebox{45}{$\sss m^2=4\L/3 ,\;\;\mbox{\tiny DoF = 4}$}}
\put(268,58){\rotatebox{14}{$\sss m^2=2\L ,\;\;\mbox{\tiny DoF = 6}$}}

\put(265,32){\tiny DoF = 7}
\put(265,22){\tiny UNITARY}
\put(284,82){\tiny NON-UNITARY}
\put(284,74){\tiny DoF=7}
\put(248,107){\tiny NON-UNITARY}
\put(248,94){\tiny DoF=7}

\put(243,50){\vector(1,0){85}}
\put(243,5){\vector(0,1){92}}

\thicklines

\put(245,5){\line(0,1){87}}
\put(245,50){\line(4,1){85}}
\put(245,50){\line(1,1){60}}

\end{picture}
\ee
The field equations are
\bea
{\cal
G}_{\m\n\r}&=&(D^2-m^2+\L/3)\,\phi_{\m\n\r}-3D_{(\m}D.\phi_{\n\r)}+
3D_{(\m}D_{\n}\phi_{\r)\s}{}^\s\nn\\
&-&3g_{(\m\n}\,\Big(
(D^2-m^2+2\L)\,\phi_{\r)\s}{}^\s\nn
-D.D.\phi_{\r)\s}{}^\s
+\frac{1}{2}\,D_{\r)} D.\phi_\s{}^\s\Big)\nn\\
&+&m\,g_{(\m\n} D_{\r)}\,\chi
\,= 0\,  ,
\label{phi3}\\
{\cal G}_5&=&-2\,(D^2-4m^2+8\L)\,\chi-m\,D.\phi_\s{}^\s\, = 0\, .
\label{chi}
\eea
and the auxiliary  $\chi$ decouples in the strictly
massless limit $m=0$.

Here we find three new gauge invariances
\bea
\!\!\!\!\!\!\delta \phi_{\m\n\r}=D_{(\m}D_{\{\n}D_{\r\})}\xi
+\frac{\L}{2}\,g_{(\m\n}D_{\r)}\xi\; ,
\!\!\!&\!\!\!&\!
\delta \chi=-\frac{1}{2}\sqrt{\frac{\L}{2}}\,(D^2+\frac{10\L}{3})\xi
\label{fader}\\
\delta \phi_{\m\n\r}=D_{(\m}D_{\{\n}
\xi_{\r\})}+\frac{\L}{3}\,g_{(\m\n}\xi_{\r)}\; ,
\!\!\!&\!\!\!&\!
\delta \chi=-\frac{1}{2}\,\sqrt{\frac{\L}{3}}\,D.\xi
\label{sun}\\
\delta \phi_{\m\n\r}=D_{(\m}\xi_{\{\n\r\})}\; ,
\!\!\!&\!\!\!&\!
\delta\chi=0\; 
\label{holeyghost}
\eea
along lines
\be
m^2-2\L=0\; ,\qquad m^2-4\L/3=0\;,  \qquad m^2=0\,, 
\ee
respectively. Again the accompanying Bianchi identities become
the massive constraints
ensuring
$7=2s+1$ PDoF for detuned $(m^2,\L)$.

The factors $m^2-2\L$, $m^2-4\L/3$ and $m^2=0$ control positivity of
lower helicity equal time commutators and therefore unitarity:
Starting from the unitary Minkowski region $m^2>2\L$ we first encounter
the gauge
line $m^2=2\L$. This is a unitary partially massless dS theory of helicities
$(\pm3,\pm2,\pm1)$ since the invariance~\eqn{fader} 
excises the lowest helicity
0 state, which reemerges as a ghost in the non-unitary region 
$4\L/3<m^2<2\L$.
At the next gauge line
$m^2=4\L/3$, the vector invariance~\eqn{sun} removes the dangerous 
helicities $(0,\pm1)$
whose commutators {\it both} flip sign upon entering the final
forbidden region $m^2<4\L/3$ where helicities $\pm1$ are now 
ghostlike. The boundary at $m^2=0$ is the strictly massless unitary 
theory where only
helicities $\pm3$ remain thanks to traceless-symmetric
invariance~\eqn{holeyghost}.    


Simple counting arguments 
imply that novel gauge invariances and partially
massless theories cutting the $(m^2,\L)$ plane into unitarily allowed
and forbidden regions are generic:
Define 
the number of components $(s)$ of a symmetric $s$-index tensor 
$\phi_{\m_1\ldots\m_s}$ as
$
\sy s\equiv
(s+1)(s+2)(s+3)/3!$ ($(s)\equiv0$ for $s<0$) as well as
$\syt s=\sy s-\sy{s-2}$ ($=(s+1)^2$ for $s\geq0$) 
when requiring tracelessness and
$\sytt s=\sy s- \sy{s-4}$ ($=2(s^2+1)$ for $s\geq1$)
for double-tracelessness.


First consider massless bosons.
The field content is a double-traceless $s$ index symmetric tensor
and the $s-1$ index gauge parameter is traceless:
\begin{center}
\begin{tabular}{|rl|c|}
\hline
&Fields&$\ss \sytt s$ \\ \hline \hline 
$\!-\hspace{-.3cm}$ &Gauge&$\ss 2.\syt{s-1}$\\
\hline
\end{tabular}
\end{center}
{\it i.e.}, $2\syt{s-1}$ DoF can be gauged away and the PDoF are  
$\sytt s-2\syt{s-1}=2$ ($s\geq1$),
helicities $\pm s$.


For massive theories, the massless field components are augmented, for $s>2$, 
by traceless
symmetric auxiliary fields $(\chi,\ldots,\chi_{\m_1\ldots\m_{s-3}})$.
Each divergence of the 
$s$-index symmetric field equations is a constraint when the remaining
open 
indices are
traceless:
\begin{center}
\begin{tabular}{|rl|cl|}
\hline
&Fields&$\ss \sytt s$&\\ 
$\!+\hspace{-.3cm}$ &Auxiliaries&$\ss \syt 0+\cdots+\syt {s-3}$&\\ 
\hline\hline
$\!-\hspace{-.3cm}$ &Constraints&$\ss\syt 0+\cdots+\syt {s-3}$&
$\hspace{-.37cm}\ss+\,\syt{s-2}+\syt{s-1}$
\\\hline
\end{tabular}
\end{center}
and PDoF are $\sytt s-\syt{s-2}-\syt{s-1}=2s+1$, the sum
of all helicities.


For partially massless higher spin theories, 
there are as many possible gauge lines as divergences of
the $s$-index symmetric field equations. On a line where a constraint
with $t$ divergences becomes a Bianchi identity we have:
\begin{center}
\begin{tabular}{|rl|cl|}
\hline
&Fields&$\ss \sytt s$&\\ 
$\!+\hspace{-.3cm}$ & Auxiliaries&
$\ss \syt 0+\cdots+\syt{s-t}+\syt{s-t+1}
+\cdots+\syt {s-3}$
&\\
\hline\hline
$\!-\hspace{-.3cm}$ & Constraints&
$\ss \phantom{\syt 0 +\cdots +2.\syt{s-t}+}
\!\!\!\!\syt{s-t+1}
+\cdots+\syt {s-3}$ &
$\!\!\!\!\!\!\ss +\syt{s-2}+\syt{s-1}$\\
$\!-\hspace{-.3cm}$ & Gauge&
$ 
\hspace{-1.8cm}\ss2.\syt{s-t}$&\\\hline
\end{tabular}
\end{center}
There are therefore
$
2t+\syt 0 +\cdots + \syt {s-2-t}\, 
$
PDoF.
When the new Bianchi identity is the
scalar one with the maximal $t=s$ divergences, there are $2s$ PDoF, 
the helicity $0$ mode having been removed. For the vector Bianchi with
$t=s-1$ there are $2s-2$ PDoF and helicities $(0,\pm1)$ are excised.
For even lower values of $t$, the sum $\syt 0 +\cdots + \syt {s-2-t}$
is non-empty and corresponds to leftover auxiliary fields.
The strictly massless case $t=1$ yields $2$ PDoF along with 
decoupled but propagating auxiliary fields 
$\syt 0 +\cdots + \syt {s-3}$, as exemplified by the
$s=3$ scalar auxiliary $\chi$.

A similar counting applies for half integer spins:
successive lower helicities are removed along gauge
lines, modulo leftover auxiliaries.
  

Having found that all massive $s>1/2$ fields in (A)dS become partially
massless along lines in the $(m^2,\L)$ plane, a pressing question is their
precise location for $s>3$: this determines the unitarity of the new
partially massless gauge theories.  
For bosons, in contrast to the first order fermions, 
covariant derivatives in their field equations
can be ordered so that the strictly massless line remains at $m^2=0$.
This suggests that the unitary and non-unitary behaviors
found in the partially massless $s=3$ and $s=5/2$ theories, respectively, 
exemplify a general distinction between $s>3$ bosons and fermions.
A large body of research exists for massless (A)dS higher
spins (see~\cite{Brink:2000ag} and references therein). The study of these
theories in $d=5$ and their Kaluza--Klein reduction to $d=4$ ought
be relevant to the new partially massless theories described in this Letter.
Whether the pattern uncovered here can be traced to a more fundamental system's
expansion remains to be seen.

We thank I. Bengtsson, A. Higuchi and R. Woodard for communications.
This work was supported by NSF grant PHY99-73935.


\newpage

\begin{thebibliography}{99} 

\bibitem{Deser:1983tm}
S.~Deser and R.~I.~Nepomechie,
Phys.\ Lett.\ B {\bf 132}, 321 (1983);
Annals Phys.\ {\bf 154}, 396 (1984).

\bibitem{Johnson:1961vt}
K.~Johnson and E.~C.~Sudarshan,
Annals Phys.\ {\bf 13}, 126 (1961).

\bibitem{Velo:1969bt}
G.~Velo and D.~Zwanziger,
Phys.\ Rev.\ {\bf 186}, 1337 (1969).

\bibitem{Higuchi:1987py}
A.~Higuchi,
Nucl.\ Phys.\ B {\bf 282}, 397 (1987);
{\it ibid} {\bf 325}, 745 (1989);
J.\ Math.\ Phys.\ {\bf 28}, 1553 (1987).


\bibitem{Townsend:1977qa}
P.~K.~Townsend,
Phys.\ Rev.\ D {\bf 15}, 2802 (1977).

\bibitem{Grassi:2001dm}
P.~A.~Grassi and P.~van Nieuwenhuizen,
Phys.\ Lett.\ B {\bf 499}, 174 (2001)
[hep-th/0011278].

\bibitem{Deser:2000de}
S.~Deser and A.~Waldron,
Phys.\ Lett.\ B,  {\bf 501}, 134 (2001)
[hep-th/0012014].

\bibitem{Pilch:1985aw}
K.~Pilch, P.~van Nieuwenhuizen and M.~F.~Sohnius,
Commun.\ Math.\ Phys.\ {\bf 98}, 105 (1985).

\bibitem{Deser:2001}
S.\ Deser and A.\ Waldron, to appear.

\bibitem{Singh:1974rc}
L.~P.~Singh and C.~R.~Hagen,
Phys.\ Rev.\ D {\bf 9}, 910 (1974);
{\it ibid} 898 (1974).
F.~A.~Berends, J.~W.~van Holten, P.~van Nieuwenhuizen and B.~de Wit,
Nucl.\ Phys.\ B {\bf 154}, 261 (1979).

\bibitem{Aragone:1987yx}
C.~Aragone, S.~Deser and Z.~Yang,
Annals Phys.\ {\bf 179}, 76 (1987).

\bibitem{Fronsdal:1978rb}
C.~Fronsdal,
Phys.\ Rev.\ D {\bf 18}, 3624 (1978);
J.~Fang and C.~Fronsdal,
Phys.\ Rev.\ D {\bf 18}, 3630 (1978);
T.~Curtright,
Phys.\ Lett.\ B {\bf 85}, 219 (1979);
B.~de Wit and D.~Z.~Freedman,
Phys.\ Rev.\ D {\bf 21}, 358 (1980).

\bibitem{Brink:2000ag}
L.~Brink, R.~R.~Metsaev and M.~A.~Vasiliev,
Nucl.\ Phys.\ B {\bf 586}, 183 (2000)
[hep-th/0005136].

\end{thebibliography}
\end{document}